%% file: charm13.tex
\documentclass[12pt]{article}
\usepackage{graphicx}
\usepackage{bm}

\def\pbnr{}
\def\speaker{Eric Braaten}

\def\title{Theoretical Interpretations of the $XYZ$ Mesons}
\def\affiliation{Physics Department, The Ohio State University\\
 Columbus, Ohio, USA}
\def\support{This research was supported in part by the U.S.\ Department of Energy
under grant DE-FG02-91-ER40690.}

\input charmmacros.tex

\begin{document}
\begin{titlepage}
\pubblock

\vfill
\Title{\title}
\vfill
\Author{\speaker\SupportedBy{\support}}
\Address{\affiliation}
\vfill
\begin{Abstract}
The $XYZ$ mesons are unexpected mesons containing 
a heavy quark-antiquark pair that have been discovered during the last decade.  
The models for the $XYZ$ mesons  that have been proposed 
include conventional quarkonium, quarkonium hybrids, 
and quarkonium tetraquarks, whose four constituents can be clustered in several possible ways.
None of the models have provided a compelling pattern for the $XYZ$ mesons.
The most promising theoretical approaches within QCD 
are lattice QCD for the $c \bar c$ mesons,
lattice NRQCD for the $b \bar b$ mesons,
and the Born-Oppenheimer approximation.
The additional hints that will be provided by future experiments 
guarantee the eventual solution of the $XYZ$ puzzle. 
\end{Abstract}
\vfill
\begin{Presented}
\venue
\end{Presented}
\vfill
\end{titlepage}
\def\thefootnote{\fnsymbol{footnote}}
\setcounter{footnote}{0}
%

\section{The $\bm{XYZ}$ Mesons}
\label{sec:XYZ}

The $XYZ$ mesons are unexpected mesons 
discovered during the last decade that contain 
a heavy quark-antiquark pair and are above the open-heavy-flavor threshold.  
Some of the more surprising of these $XYZ$ mesons were
\begin{itemize}
\item
$X(3872)$, 
discovered by the Belle Collaboration in 2003 \cite{Choi:2003ue}.  
It has comparable branching fractions into $J/\psi\, \rho$ 
and $J/\psi\, \omega$, implying a severe violation of isospin symmetry.
\item
$Y(4260)$, 
discovered by the BaBar Collaboration in 2005 \cite{Aubert:2005rm}.  
It has $J^{PC}$ quantum numbers $1^{--}$, but it is produced very weakly 
in $e^+e^-$ annihilation.
\item
$Z_b^+(10610)$ and $Z_b^+(10650)$, 
discovered by the Belle Collaboration in 2011 \cite{Belle:2011aa}.  
They both decay into $\Upsilon\, \pi^+$, 
revealing that they are tetraquark mesons with constituents 
$b\bar b u \bar d$.
\item
$Z_c^+(3900)$, 
discovered by the BESIII Collaboration in 2013 \cite{Ablikim:2013mio}.  
It decays into $J/\psi\, \pi^+$, revealing that it is a tetraquark meson 
with constituents $c\bar c u \bar d$.
\end{itemize}
Updated lists of the $XYZ$ mesons in both the $c \bar c$ and $b \bar b$
sectors are given in Ref.~\cite{Bodwin:2013nua}.
The list of new $c \bar c$ mesons above the 
$D \bar D$ threshold includes 15 neutral and 5 charged mesons.  
The list of new $b \bar b$ mesons above the $B \bar B$ threshold 
includes 1 neutral and 2 charged mesons.  
Many of these mesons are surprisingly narrow.  
They reveal a serious gap in our understanding of the QCD spectrum.

The $XYZ$ puzzle is the problem of understanding the nature of these new mesons.
Most of the theoretical work on the $XYZ$ mesons during the last decade 
has been carried out using simple-minded models 
that make no direct contact with QCD.  
The models that have been proposed are summarized in Section \ref{sec:models}.  
The most promising theoretical approaches to the $XYZ$ mesons 
within QCD all involve lattice gauge theory and
are discussed in Section \ref{sec:QCD}.  
The extensive information on the $XYZ$ mesons that is already available
and the additional hints that will 
be provided by future experiments almost guarantee 
the eventual solution of the $XYZ$ puzzle.

\section{Constituent Models}
\label{sec:models}

Most of the theoretical work on the $XYZ$ mesons has been carried out  
using models that may be compatible with QCD but make no direct contact 
with the fundamental field theory.  The models
can be classified according to their constituents.  
The basic categories are (1) quarkonium, whose constituents 
are a heavy quark and a heavy antiquark, 
(2) quarkonium hybrid, which has an additional gluonic constituent,
 and (3) quarkonium tetraquark, 
whose additional constituents are a light quark and antiquark.  
The tetraquark category can be subdivided into further categories 
according to the way the four constituents are clustered inside the meson.  
None of the models that has been proposed thus far has provided a 
compelling pattern for the $XYZ$ mesons that have been observed.

\subsection{Conventional quarkonium}
\label{sec:onium}

The constituents of a conventional quarkonium consist of a 
heavy quark $Q$ and a heavy antiquark $\bar Q$ only.  
There is a well-developed phenomenology for conventional quarkonium 
based on potential models. It provides an accurate description 
of the quarkonium states below the open-heavy-flavor threshold.  
Its accuracy for states above the open-heavy-flavor has not been tested, 
but it gives fairly well-defined predictions for their masses 
and for radiative transition rates to other quarkonium states \cite{Barnes:2005pb,Eichten:2007qx}.  
The predicted states form
orbital-angular-momentum multiplets of states 
related by heavy-quark spin symmetry:  $S$-wave multiplets 
with $J^{PC}$ quantum numbers $\{0^{-+},1^{--}\}$, 
$P$-wave multiplets $\{1^{+-}, (0,1,2)^{++} \}$, 
$D$-wave multiplet $\{2^{-+}, (1,2,3)^{--}\}$, etc.

\subsection{Quarkonium hybrid}
\label{sec:hybrid}

A quarkonium hybrid consists of $Q$, $\bar Q$, and a gluonic excitation $g$.  
There are two important model-independent 
properties of a quarkonium bybrid:
\begin{itemize}
\item
The wave function for the $Q\bar Q$ pair near the origin is very small.  
This follows from the fact that when the $Q$ and $\bar Q$ 
pair are 
close together, they must be in a color-octet state 
in which the $Q$ and $\bar Q$ have a repulsive $1/R$ potential 
at short distance. 
\item
Decays into two S-wave mesons are suppressed, so decays into an 
$S$-wave and a P-wave meson are preferred, provided they are 
kinematically accessible \cite{Close:1994hc,Kou:2005gt}.
\end{itemize}

There are two rather different physical pictures of a quarkonium hybrid.  
In the constituent-gluon picture, the excitation of the gluon field is 
interpreted as a particle with definite $J^{PC}$ quantum numbers.  
The lowest energy levels of a quarkonium hybrid can be reproduced 
by taking the constituent gluon to have $J^{PC}=1^{+-}$, 
which can be interpreted as a vector particle in a $P$-wave orbital.  
If the $Q \bar Q$ pair is in an S-wave state, 
the resulting spin-symmetry multiplet for the hybrid is 
$\{1^{--}, (0,1,2)^{-+}\}$.  If the $Q \bar Q$ pair is in a $P$-wave state, 
there are three spin-symmetry multiplets: $\{0^{++}, 1^{+-}\}$, 
$\{1^{++}, (0,1,2)^{+-}\}$, and $\{2^{++}, (1,2,3)^{+-}\}$.  
The $1^{-+}$, $0^{+-}$, and $2^{+-}$ states have exotic quantum numbers
that are not possible if the only constituents are $Q$ and $\bar Q$.  

There is an alternative picture of a quarkonium hybrid 
that can be called a {\it Born-Oppenheimer hybrid}.  
The excitation of the gluon field is a configuration of 
chromoelectric and chromomagnetic fields in which the 
$Q$ and $\bar Q$ are embedded and for which they provide the source.  
The gluon field energy provides a potential that determines the 
motion of the $Q$ and $\bar Q$ and the gluon field responds 
almost instantaneously to their motion.  
The possible gluon field configurations are traditionally labelled 
by upper-case Greek letters with subscripts and superscripts, 
such as $\Pi_u$, $\Sigma_u^-$, etc.  
The lowest energy level of the hybrid is a $P$-wave $Q \bar Q$ pair 
in the $\Pi_u$ field, which consists of spin-symmetry multiplets 
$\{1^{--}, (0,1,2)^{-+}\}$ and $\{1^{++}, (0,1,2)^{+-}\}$.  
Another low-lying energy level 
is an S-wave $Q\bar Q$ pair in the $\Sigma_u^-$ field, 
which gives a single spin-symmetry multiplet 
$\{0^{++},1^{+-} \}$.  Note that the same spin-symmetry multiplets 
are obtained as in the constituent gluon picture, 
but they arise in different ways.	

There is an  $XYZ$ state that is a compelling candidate 
for a charmonium hybrid: the $Y(4260)$.  
It has quantum numbers $1^{--}$, which is one of the quantum numbers 
in the ground-state spin-symmetry multiplet.  
Despite these quantum numbers, it is produced very weakly 
in $e^+e^-$ annihilation: the resonance appears as a small peak 
near a deep minimum in the cross section 
for $e^+e^-$ annihilation into hadrons. 
This small production rate in $e^+e^-$ annihilation indicates 
that the wave function for the $c \bar c$ pair at the origin is 
very small, which is a characteristic property of a quarkonium hybrid.  
The decay of the $Y(4260)$ into the charm mesons $D \bar D$  
has not been observed in spite of the large available phase space.  
This is consistent with the suppression of decays of a hybrid into 
a pair of $S$-wave mesons.  A convincing case for the $Y(4260)$ as a 
charmonium hybrid was made by Close and Page in 2005, 
shortly after its discovery \cite{Close:2005iz}.  The recent discovery
of the unexpected $Z_c\, \pi$ decay mode 
of the $Y(4260)$ by the BESIII Collaboration   \cite{Ablikim:2013mio} 
has not made the case any less compelling.

\subsection{Compact tetraquark}
\label{sec:compact}

A quarkonium tetraquark whose four constituents $Q \bar Q q \bar q$ 
all have overlapping spatial wavefunctions is called a {\it compact tetraquark}. 
A simple model for their interaction is the quark potential model 
in which each pair of constituents interacts through a potential.  
Careful solutions of the 4-body problem reveal that 
the tetraquark is unstable with respect to fall-apart decays into 
two mesons unless the mass of the tetraquark is below the thresholds 
for all pairs of mesons with the appropriate quantum numbers, 
both a pair of heavy-light mesons $Q \bar q$ and $ \bar Q q$ 
and also a heavy quarkonium $Q \bar Q$ 
plus a light meson $q \bar q$ \cite{Vijande:2007rf}. 
Since the $XYZ$ mesons are above the threshold for $Q \bar q$ 
and $\bar Q q$ mesons, this implies that they cannot be compact 
tetraquarks unless the interactions between the constituents 
are more complicated than pairwise potentials, involving, 
for example, 3-body and 4-body potentials.
 
\subsection{Meson molecule}
\label{sec:molecule}

A quarkonium tetraquark whose substructure consists of a pair of mesons, 
which are color-singlet clusters with constituents $Q \bar q$ and $\bar Q q$,
is called a {\it meson molecule}.  
In order to be a plausible constituent for a molecule, 
a meson must be rather narrow.  
The nonstrange charm mesons that are plausible constituents are 
$D$, $D^*$, $D^*_0$, and $D_1$.  The strange charm mesons 
that are plausible constituents are $D_s$, $D^*_s$, $D^*_{s0}$, 
$D_{s1}$, and $D^*_{s2}$.

One of the motivations for meson molecules is that many of the $XYZ$ mesons 
have mass near a threshold for a pair of heavy-light mesons.  
This could be a coincidence or it could be a consequence of 
attractive interactions between the mesons that create the molecule.  
Since there are 25 thresholds between 3770~MeV and 5150~MeV  
for pairs of mesons with no net strangeness,
a randomly chosen mass in this region is likely 
to be within 50~MeV of some threshold.
It is therefore not implausible that the proximity of the masses to thresholds 
is coincidental. 

The possibility that charm mesons could be bound into molecules 
by their interactions was first studied quantitatively by Tornqvist 
in 1995 \cite{Tornqvist:1993ng}.  He considered pion-exchange interactions 
between the charm mesons with an ultraviolet cutoff to regularize a divergence 
from short distances.  This model predicts no molecules 
in isospin-1 channels, but it predicts bound states near threshold 
in several isospin-0 $J^{PC}$ channels: $D^* \bar D$ with $0^{-+}$ 
and $1^{++}$ and $D^* \bar D^*$ with $0^{++}$, $0^{-+}$, $1^{+-}$, 
and $2^{++}$.  Whether or not the molecule is actually bound, 
with mass below the threshold, depends on the ultraviolet cutoff.
The binding energy is also sensitive to additional interactions 
between the charm mesons, such as the potential from the exchange 
of $\rho$, $\omega$, and other mesons.  In the case of bottom mesons, 
the binding energies are larger than for charm mesons by about 50~MeV.
The meson pairs $B^* \bar B$ and $B^* \bar B^*$ should therefore be bound
in all the $J^{PC}$ channels listed above.  
However, the more strongly the mesons are bound, the more likely 
it is that QCD interactions rearrange their constituents in a way
that is not well-approximated by a pair of mesons. 

There is one $XYZ$ meson whose identification as a meson molecule 
is completely unambiguous and this is the $X(3872)$.  
This follows from the universality of S-wave near-threshold resonances
\cite{Braaten:2003he}.  If a resonance is close enough to a threshold 
and if it has quantum numbers that allow an $S$-wave coupling 
to the threshold, it is transformed by its interactions 
into a weakly-bound molecule with universal properties 
that are determined only by its binding energy $E_b$.  
One such property is the mean-square separation of its constituents 
$\langle r^2 \rangle = 1/(4\mu E_b)$, where $\mu$ is the 
reduced mass of the constituents.  The universal behavior 
requires the state to be very close to threshold, 
within 10~MeV for charm mesons and within 3~MeV for bottom mesons.  
This criterion is easily satisfied by the $X(3872)$. 
Precise measurements of its mass by the CDF and Belle Collaborations 
imply that it is below the $D^{*0}\bar D^0$ threshold by 
$0.3 \pm 0.4$~MeV \cite{Aaltonen:2009vj,Choi:2011fc}.  
The quantum numbers  of the $X(3872)$ have recently been 
determined definitively by the LHC Collaboration to be $1^{++}$ \cite{Aaij:2013zoa}.  
This implies that it has an S-wave coupling to $D^{*0} \bar D^0$.  
Thus, regardless of the source of the attraction between the charm mesons, 
the $X(3872)$ must be a loosely bound charm meson molecule 
whose constituents are a superposition of $D^{*0} \bar D^0$ 
and $D^0 \bar D^{*0}$.  The universal prediction for the rms 
separation of the charm mesons is about 5~fm 
if the binding energy is 0.3~MeV.  Thus the $X(3872)$ has a truly
astonishing spatial extent that
is probably an order of magnitude larger than that of most hadrons.

Molecular interpretations have been proposed for most of the $XYZ$ mesons.  
In most cases, the identification as a molecule has not proved to be very 
predictive.  One exception is the $Y(4260)$, which has been proposed as a 
$D_1 \bar D$ molecule \cite{Ding:2008gr,Wang:2013cya}.
Another exception is the $Z_b(10610)$ and $Z_b(10650)$, 
which have been proposed as $B^* \bar B$ and $B^* \bar B^*$ molecules,
respectively \cite{Bondar:2011ev}.

\subsection{Diquark-onium}
\label{sec:diquark}

A quarkonium tetraquark whose substructure consists of diquarks,
which are colored clusters $Qq$ and $\bar Q \bar q$,
is called {\it diquark-onium} \cite{Drenska:2010kg}.  
The diquark $Qq$ is usually treated as a particle with anti-triplet color
whose total spin can be 0 or 1.  The spectrum of diquark-onium
is then determined using purely algebraic methods,
like in the naive quark model for mesons and baryons.
If we only consider the 
light quarks $u$ and $d$ and if we take the diquarks to be in an 
$S$-wave state, the predicted diquark-onium states are degenerate 
isospin-0 and isospin-1 multiplets for each of six $J^{PC}$ states: 
two with $0^{++}$, two with $1^{++}$, one with $1^{+-}$, and one with $2^{++}$.  
We can also allow the light quark to be $s$, 
and we can consider orbital angular momentum excitations 
and radial excitations of the diquark pair.  
Since there is no dynamics limiting the possible quantum numbers,
the result is a proliferation of predicted diquark-onium states.

\subsection{Hadro-quarkonium}
\label{sec:hadro}

A quarkonium tetraquark whose substructure consists of a
compact color-singlet $Q \bar Q$ pair to which a light $q \bar q$ 
pair is bound is called {\it hadro-quarkonium} \cite{Dubynskiy:2008mq}. 
Alternatively, it can be regarded as a light $q \bar q$ meson
bound to a quarkonium.  
The motivation for hadro-quarkonium is that many of the $XYZ$ mesons
have been observed only through a single hadronic transition 
to a quarkonium and a light meson, such as $J/\psi\, \omega$.  
Hadro-quarkonium provides a simple-minded explanation for this fact 
by essentially assuming that the quarkonium and the light meson 
are already preexisting in the $XYZ$ meson.

\subsection{Born-Oppenheimer tetraquark}
\label{sec:BOtetra}

A new proposal for the structure of a quarkonium tetraquark 
that was recently introduced  is a
{\it Born-Oppenheimer tetraquark}  \cite{Braaten:2013boa}.  
It is motivated by the 
Born-Oppenheimer picture for a quarkonium hybrid, in which the $Q$ 
and $\bar Q$ are embedded in a gluon field configuration for which 
they provide the source.  
The field configuration in a quarkonium hybrid could also include 
light-quark fields, but it must be a flavor singlet. The proposal  
for Born-Oppenheimer tetraquarks is based on the observation 
that the configuration of gluon and light-quark fields 
in which the $Q$ and $\bar Q$ are embedded need not be a flavor singlet. 
It could instead have isospin 1, in which case the meson is a tetraquark. 
The color and spatial structure of a Born-Oppenheimer tetraquark 
is qualitatively different from each of the possibilities discussed 
in the previous four sections. It will be discussed in more detail 
in Section \ref{sec:BOpot}.

\section{Theoretical approaches within QCD}
\label{sec:QCD}

A deep understanding of the $XYZ$ mesons must eventually be based 
on the fundamental field theory of QCD. 
The most promising approaches to this problem all involve 
lattice gauge theory in an essential way. They are 
(1) lattice QCD, which can be applied directly to $c \bar c$ mesons, 
(2) lattice NRQCD, which can be applied to $b\bar b$ meson, and 
(3) the Born-Oppenheimer approximation, which can be applied 
to both $c \bar c$ and $b\bar b$ mesons.

\subsection {Lattice QCD for $\bm{c \bar c}$ mesons}
\label{sec:latticeQCD}

The mass of the charm quark is small enough that lattice QCD 
can be applied directly to $c \bar c$ mesons
using the computational resources that are currently available. 
Dudek, Edwards, Mathur, and Richards carried out pioneering calculations 
of the $c \bar c$ meson spectrum that included many states above the 
$D\bar D$ threshold \cite{Dudek:2007wv}. These calculations have been 
extended by the Hadron Spectrum Collaboration \cite{Liu:2012ze}. 
They used a lattice with 
$24^3\times 128$ sites and a spatial lattice spacing of 0.12~fm. 
They had dynamical light quarks, but the masses of the $u$ and $d$ quarks 
were unphysically heavy, corresponding to a pion mass of 400~MeV. 
Their results were not definitive, because they did not carry out 
the extrapolations to 0 lattice spacing and to the physical 
$u$ and $d$ quark masses that are necessary to quantify all 
systematic errors. Nevertheless their results are very impressive. 
They used the cross correlators of many operators to determine 
the spectrum of flavor-singlet $c\bar c$ mesons. They were able to 
identify 46 statistically significant states in various $J^{PC}$ channels, 
with spins as high as 4 and with masses as high as 4.6~GeV. 
They also demonstrated that they were able to discriminate between 
charmonium and charmonium hybrids based on the strengths with which they 
couple to various operators. Their charmonium states filled out 
complete $1S$, $2S$, $3S$, $1P$, $2P$, $1D$, and $1F$ multiplets. 
The lowest charmonium hybrids formed the spin-symmetry multiplet 
$\{1^{--},(0,1,2)^{-+}\}$. The higher charmonium hybrids filled out 
three additional multiplets: $\{1^{++},(0,1,2)^{+-}\}$, 
$\{0^{++}, 1^{+-} \}$, and $\{2^{++}, (1,2,3)^{+-}\}$. 
The quantum numbers $1^{-+}$, $0^{+-}$, and $2^{+-}$ are exotic.

The ground-state charmonium hybrid includes a $1^{--}$ state 
that can be identified with the $Y(4260)$. 
The results for the masses of charmonium hybrids in Ref.~\cite{Liu:2012ze} 
are not definitive, because they were not extrapolated to 
zero lattice spacing or to the physical $u$ and $d$ quark masses. 
It is plausible that the mass splitting within spin-symmetry multiplets 
are less sensitive to these 
extrapolations than the masses themselves. If this is the case, 
we can get better predictions for the masses of charmonium hybrids 
in the ground-state multiplet by taking the mass of the $Y(4260)$ 
from experiment and the mass splittings from Ref.~\cite{Liu:2012ze}. 
The resulting predictions for the masses of the $0^{-+}$, $1^{-+}$, 
and $2^{-+}$ charmonium hybrids are $4173\pm 21$~MeV, 
$4195\pm 23$~MeV, and $4312\pm 24$~MeV, respectively \cite{Braaten:2013boa}. 
It would of course be preferable to have definitive lattice QCD calculations 
of the hybrid charmonium masses which have been extrapolated to zero lattice spacing 
and to the physical masses of the light quarks.

\subsection{Lattice NRQCD for $\bm{b \bar b}$ mesons}
\label{sec:latticeNR}

The mass of the bottom quark is too large to apply lattice QCD
directly to $b \bar b$ mesons
with currently available computational resources.
Instead it is necessary to use an effective field theory 
called {\it NonRelativistic QCD} (NRQCD)
in which the $b$ quark is treated nonrelativistically.  
Lattice NRQCD was used by Juge, Kuti, and Morningstar 
to calculate the masses of bottomonium hybrid \cite{Juge:1999ie}.  
They used a lattice with $15^3 \times 45$ sites 
and a spatial lattice spacing of 0.115~fm. 
They had no dynamical light quarks, so their results are only
useful qualitatively. Their ground-state charmonium hybrid 
was the spin-symmetry multiplet $\{1^{--}, (0,1,2)^{-+}\}$.  
The next few multiplets were $\{1^{++}, (0,1,2)^{+-}\}$, 
$\{0^{++}, 1^{+-}\}$, and then a radially excited 
$\{1^{--}, (0, 1, 2)^{-+}\}$ multiplet.

\subsection{Born-Oppenheimer approximation}
\label{sec:BOpot}

Another approach to the $XYZ$ mesons within QCD is based 
on the Born-Oppenheimer approximation.  
It provides a unified framework for describing conventional quarkonium,
quarkonium hybrids, and quarkonium tetraquarks.  
The Born-Oppenheimer approximation is used in atomic physics 
to describe molecules.  A Born-Oppenheimer potential
is the energy of the electrons  
in the presence of static sources for the atomic nuclei 
separated by a distance $R$, and it can be interpreted as 
the potential energy of the nuclei. 
In the Born-Oppenheimer approximation, the motion of the nuclei 
is described by the Schroedinger equation in the 
Born-Oppenheimer potential, while the much lighter electrons 
are assumed to respond 
instantaneously to the motion of the nuclei.  

The Born-Oppenheimer approximation for QCD applied to hadrons 
containing a heavy quark and antiquark was developed by 
Juge, Kuti, and Morningstar \cite{Juge:1999ie}.
A Born-Oppenheimer (B-O) 
potential can be defined by the energy of gluon and 
light-quark fields in the presence of static $Q$ and $\bar Q$ 
sources separated by a distance $R$, and it can be interpreted as 
the potential energy of the $Q \bar Q$ pair.  
In the Born-Oppenheimer approximation, the motion of the $Q$ and $\bar Q$ 
is described by the Schroedinger equation in the B-O  potential, while
the gluon and light-quark fields are assumed to respond instantaneously 
to the motion of the $Q$ and $\bar Q$.  The possible configurations 
of gluon and light-quark fields can be labelled by quantum numbers 
of symmetry operators in the presence of the $Q$ and $\bar Q$ sources.
Juge, Kuti, and Morningstar followed the tradition 
of atomic physics of labeling them instead by upper-case Greek letters 
with a subscript and possibly a superscript: 
$\Sigma_g^+$, $\Pi_u$, $\Sigma_u^-$, etc.  
They used lattice QCD without dynamical quarks to calculate the lowest 
few B-O potentials.  Their lattice had only
$10^3 \times 30$ sites and a spatial lattice spacing of 0.1~fm.  
The ground-state potential $\Sigma_g^+$ can be identified
with the potential used in quark models,
which is attractive and proportional to $1/R$ 
at short distances and repulsive and linear in $R$ at long distances.  
The first excited potential $\Pi_u$ and the second excited potential
$\Sigma_u^-$ have minima for $R$ near 0.3~fm and 0.15~fm, respectively.  
They are both repulsive and linear in $R$ at large $R$. At small $R$, 
they both approach the repulsive $1/R$ potential between $Q$ 
and $\bar Q$ in a color-octet state \cite{Juge:2002br}.  

The solutions to the Schrodinger equation in the B-O potentials 
can be labelled $nL$, where $n=1,2,3, \ldots$ is a radial quantum number 
and $L=0,1,2,\ldots$ (or $S$, $P$, $D$, \ldots) is an 
orbital-angular-momentum quantum number whose minimum value depends 
on the B-O potential.  The energy levels in the ground-state 
B-O potential $\Sigma_g^+$ are conventional quarkonia.  
The energy levels in the excited B-O potentials $\Pi_u$, 
$\Sigma_u^-$, etc. are quarkonium hybrids.  The lowest hybrid energy 
level is $1P$ in the $\Pi_u$ potential, which consists of the two 
spin-symmetry multiplets $\{1^{--}, (0,1,2)^{-+}\}$ and 
$\{1^{++},(0,1,2)^{+-}\}$.  In the case of bottomonium, 
the next lowest energy levels are $1S$ in the $\Sigma_u^-$ potential, 
which is a $\{0^{++}, 1^{+-}\}$ multiplet, and a $2P$ radial excitation 
in the $\Pi_u$ potential.  
These results were obtained using lattice QCD without dynamical quarks.  
Calculations of the B-O potentials with dynamical quarks would be 
required to determine the correct ordering and spacing of the 
bottomonium hybrid energy levels.  

Quarkonium tetraquarks can also be treated using the 
Born-Oppenheimer approximation.  The appropriate B-O potentials 
are energies of gluon and light-quark fields that have isospin 1 
instead of being flavor singlets.  The isospin-1 B-O potentials 
for tetraquarks can presumably be calculated using lattice QCD, 
although the presence of the light valence quarks makes the 
calculations more demanding than those for
flavor-singlet B-O potentials.  

One can make plausible guesses for the qualitative behavior 
of the isospin-1 B-O potentials \cite{Braaten:2013boa}. 
At small $R$, the isospin-1 $\Pi_u$ and $\Sigma_u^-$ potentials should 
both approach the repulsive $1/R$ potential of a color-octet 
$Q \bar Q$ pair, just like the flavor-singlet potentials except 
for an offset in the energy.  At large $R$, they should both be 
repulsive and linear in $R$.  Thus they should both have 
minima at intermediate $R$. It is therefore plausible that the 
isospin-1 $\Pi_u$ and $\Sigma_u^-$ potentials both 
have similar shapes to the corresponding flavor-singlet potentials, 
except for an offset in the energy.  The discovery of the $Z_c(3900)$ 
tetraquark with a smaller mass than the $Y(4260)$ hybrid implies 
that the isospin-1 B-O potentials are lower in energy than the 
flavor-singlet potentials.  If they have the same shapes, 
the neutral members of the lowest tetraquark multiplet should 
have the same $J^{PC}$ 
quantum numbers as the lowest hybrid multiplets.  
The lowest multiplets for charmonium tetraquarks  
in the $\Pi_u$  and $\Sigma_u^-$ potentials would then be 
$\{1^{--}, (0,1,2)^{-+}\}$, $\{1^{++}, (0,1,2)^{+-}\}$, 
and $\{0^{++}, 1^{+-}\}$.

\section{Outlook}
\label{sec:conclude}

The discoveries of the neutral $XYZ$ mesons, 
the bottomonium tetraquarks $Z_b$ and $Z_b'$, and the charmonium 
tetraquark $Z_c$ have revealed a serious gap in out understanding 
of the QCD spectrum.  None of the proposed models for the 
$XYZ$ mesons has revealed a compelling pattern.  
A promising recent proposal that has not yet been fully explored 
is that conventional quarkonia, quarkonium hybrids, and quarkonium tetraquarks 
can all be described in the same coherent framework based on the 
Born-Oppenheimer approximation.  

The most promising theoretical approaches to the $XYZ$ mesons 
within QCD all involve lattice gauge theory.  
There are several lattice calculations that would have a 
significant impact on our understanding of the $XYZ$ mesons:
\begin{itemize}
\item
definitive calculations of the spectrum of $c \bar c$ mesons using lattice QCD, 
\item
definitive calculations of the spectrum of $b \bar b$ mesons using lattice NRQCD, 
\item
calculations of the Born-Oppenheimer potentials using lattice QCD, 
both the flavor-singlet potentials for quarkonium hybrids 
and the isospin-1 potentials for quarkonium tetraquarks.  
\end{itemize}
In addition, it is important to develop a phenomenological framework 
for estimating hadronic transition rates between conventional quarkonia, quarkonium hybrids, 
and quarkonium tetraquarks.  

Many clues to the $XYZ$ puzzle have already been provided by experiment.  
It is a little embarrassing for QCD theory that this problem has remained 
unsolved  for so long.  Fortunately, many additional hints 
can be expected from ongoing experiments at BESIII, current analyses 
by the LHC collaborations, and future experiments at Belle II and Panda.  
They make the solution to the $XYZ$ puzzle almost inevitable.

\Acknowledgements
This research was supported in part by the Department of Energy 
under grant DE-FG02-91-ER40690.

\end{document}

%% file: charmmacros.tex
\newcommand\pubnumber{\pbnr}
\newcommand\pubdate{\today}
\def\Title#1{\begin{center} {\Large #1 } \end{center}}
\def\Author#1{\begin{center}{ \sc #1} \end{center}}

\newcommand{\OnBehalf}[1]{\sbox0{#1}\ifdim\wd0=0pt
        {}
	\else
	{\\on behalf of #1}
	\fi}
\newcommand{\SupportedBy}[1]{\sbox0{#1}\ifdim\wd0=0pt
        {}
	\else
	{\footnote{#1}}
	\fi}
\def\Address#1{\begin{center}{ \it #1} \end{center}}

\newcommand\pubblock{\includegraphics[width=5cm]{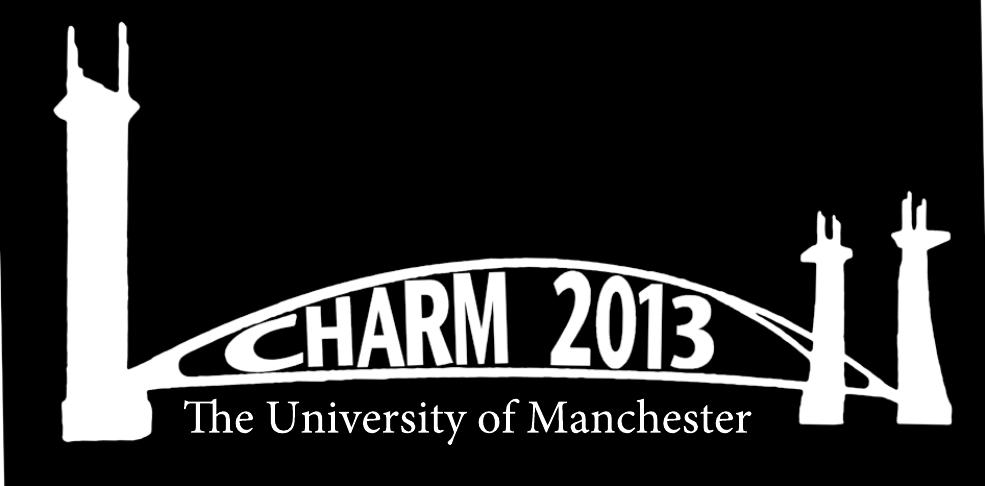}\hfill{\begin{tabular}{l} \pubnumber\\
         \pubdate  \end{tabular}}}
\newenvironment{Abstract}{\begin{quotation}  }{\end{quotation}}
\newenvironment{Presented}{\begin{quotation} \begin{center} 
             PRESENTED AT\end{center}\bigskip 
      \begin{center}\begin{large}}{\end{large}\end{center} \end{quotation}}
\def\Acknowledgements{\bigskip  \bigskip \begin{center} \begin{large}
             \bf ACKNOWLEDGEMENTS \end{large}\end{center}}
\def\venue{The 6$^{th}$ International Workshop on Charm Physics\\
(CHARM 2013)\\
Manchester, UK,  31 August -- 4 September, 2013}

\def\beq{\begin{equation}}
\def\eeq#1{\label{#1}\end{equation}}
\def\eeqn{\end{equation}}

\def\beqa{\begin{eqnarray}}
\def\eeqa#1{\label{#1}\end{eqnarray}}
\def\eeqan{\end{eqnarray}}

\let\bar=\overbar

\def\Dslash{\not{\hbox{\kern-4pt $D$}}}
\def\dslash{\not{\hbox{\kern-2pt $\del$}}}

\def\msb{{\bar{\ssstyle M \kern -1pt S}}}




%% file: charm13.bbl
\begin{thebibliography}{99}


\bibitem{Choi:2003ue}
  S.~K.~Choi {\it et al.}  [Belle Collaboration],
  Phys.\ Rev.\ Lett.\  {\bf 91} (2003) 262001
  [hep-ex/0309032].

\bibitem{Aubert:2005rm}
  B.~Aubert {\it et al.}  [BaBar Collaboration],
  Phys.\ Rev.\ Lett.\  {\bf 95} (2005) 142001
  [hep-ex/0506081].

\bibitem{Belle:2011aa}
  A.~Bondar {\it et al.}  [Belle Collaboration],
  Phys.\ Rev.\ Lett.\  {\bf 108} (2012) 122001
  [arXiv:1110.2251].

\bibitem{Ablikim:2013mio}
  M.~Ablikim {\it et al.}  [BESIII Collaboration],
  Phys.\ Rev.\ Lett.\  {\bf 110} (2013) 252001
  [arXiv:1303.5949].

\bibitem{Bodwin:2013nua}
  G.~T.~Bodwin, E.~Braaten, E.~Eichten, S.~L.~Olsen, T.~K.~Pedlar and J.~Russ,
  arXiv:1307.7425.

\bibitem{Barnes:2005pb} 
  T.~Barnes, S.~Godfrey and E.~S.~Swanson,
  Phys.\ Rev.\ D {\bf 72}  (2005) 054026
  [hep-ph/0505002].
  
\bibitem{Eichten:2007qx}
  E.~Eichten, S.~Godfrey, H.~Mahlke, and J.~L.~Rosner,
  Rev.\ Mod.\ Phys.\  {\bf 80}  (2008) 1161
  [hep-ph/0701208].

\bibitem{Close:1994hc}
  F.~E.~Close and P.~R.~Page,
  Nucl.\ Phys.\ B {\bf 443} (1995) 233
  [hep-ph/9411301].

\bibitem{Kou:2005gt}
  E.~Kou and O.~Pene,
  Phys.\ Lett.\ B {\bf 631} (2005) 164
  [hep-ph/0507119].

\bibitem{Close:2005iz}
  F.~E.~Close and P.~R.~Page,
  Phys.\ Lett.\ B {\bf 628} (2005) 215
  [hep-ph/0507199].

\bibitem{Vijande:2007rf}
  J.~Vijande, E.~Weissman, A.~Valcarce and N.~Barnea,
  Phys.\ Rev.\ D {\bf 76} (2007) 094027
  [arXiv:0710.2516].
  
\bibitem{Tornqvist:1993ng}
  N.~A.~Tornqvist,
  Z.\ Phys.\ C {\bf 61} (1994) 525
  [hep-ph/9310247].

\bibitem{Braaten:2003he}
  E.~Braaten and M.~Kusunoki,
  Phys.\ Rev.\ D {\bf 69} (2004) 074005
  [hep-ph/0311147].

\bibitem{Aaltonen:2009vj}
  T.~Aaltonen {\it et al.}  [CDF Collaboration],
  Phys.\ Rev.\ Lett.\  {\bf 103} (2009) 152001
  [arXiv:0906.5218].
  
\bibitem{Choi:2011fc}
  S.-K.~Choi, S.L.~Olsen, K.~Trabelsi, I.~Adachi, H.~Aihara, K.~Arinstein, D.M.~Asner and T.~Aushev {\it et al.},
  Phys.\ Rev.\ D {\bf 84} (2011) 052004
  [arXiv:1107.0163].

\bibitem{Aaij:2013zoa}
  R.~Aaij {\it et al.}  [LHCb Collaboration],
  Phys.\ Rev.\ Lett.\  {\bf 110} (2013) 222001
  [arXiv:1302.6269].

\bibitem{Ding:2008gr}
  G.-J.~Ding,
  Phys.\ Rev.\ D {\bf 79} (2009) 014001
  [arXiv:0809.4818].
  
\bibitem{Wang:2013cya}
  Q.~Wang, C.~Hanhart and Q.~Zhao,
  Phys.\ Rev.\ Lett.\  {\bf 111} (2013) 132003
  [arXiv:1303.6355].

\bibitem{Bondar:2011ev}
  A.~E.~Bondar, A.~Garmash, A.~I.~Milstein, R.~Mizuk and M.~B.~Voloshin,
  Phys.\ Rev.\ D {\bf 84} (2011) 054010
  [arXiv:1105.4473].

\bibitem{Drenska:2010kg}
  N.~Drenska, R.~Faccini, F.~Piccinini, A.~Polosa, F.~Renga and C.~Sabelli,
  Riv.\ Nuovo Cim.\  {\bf 033} (2010) 633
  [arXiv:1006.2741].

\bibitem{Dubynskiy:2008mq}
  S.~Dubynskiy and M.~B.~Voloshin,
  Phys.\ Lett.\ B {\bf 666} (2008) 344
  [arXiv:0803.2224].

\bibitem{Braaten:2013boa}
  E.~Braaten,
  arXiv:1305.6905.

\bibitem{Dudek:2007wv}
  J.~J.~Dudek, R.~G.~Edwards, N.~Mathur and D.~G.~Richards,
  Phys.\ Rev.\ D {\bf 77} (2008) 034501
  [arXiv:0707.4162].

\bibitem{Liu:2012ze}
  L.~Liu {\it et al.}  [Hadron Spectrum Collaboration],
  JHEP {\bf 1207} (2012) 126
  [arXiv:1204.5425].

\bibitem{Juge:1999ie}
  K.~J.~Juge, J.~Kuti and C.~J.~Morningstar,
  Phys.\ Rev.\ Lett.\  {\bf 82} (1999) 4400
  [hep-ph/9902336].

\bibitem{Juge:2002br}
  K.~J.~Juge, J.~Kuti and C.~Morningstar,
  Phys.\ Rev.\ Lett.\  {\bf 90} (2003) 161601
  [hep-lat/0207004].


\end{thebibliography}
